\documentclass[12pt,draftcls,onecolumn]{IEEEtran}

\usepackage[dvips]{color}
\usepackage{graphicx}
\usepackage{amsmath}

\begin{document}

\title{Securing Physical-Layer Communications for Cognitive Radio Networks} \normalsize

\author{Yulong~Zou,~\IEEEmembership{Senior Member,~IEEE,}
        Jia~Zhu,
        Liuqing~Yang,~\IEEEmembership{Fellow,~IEEE,}
        Ying-Chang~Liang,~\IEEEmembership{Fellow,~IEEE,} and
        Yu-Dong~Yao,~\IEEEmembership{Fellow,~IEEE}

\thanks{Y. Zou and J. Zhu are with the School of Telecommunications and Information Engineering, Nanjing University of Posts and Telecommunications, Nanjing, China. (Email: \{yulong.zou, jiazhu\}@njupt.edu.cn)}
\thanks{L. Yang is with the ECE Department, Colorado State University, Colorado State, USA. (Email: lqyang@engr.colostate.edu)}
\thanks{Y.-C. Liang is with the Institute for Infocomm Research (I2R), Agency for Science, Technology \& Research (A*STAR), Singapore. (Email: ycliang@i2r.a-star.edu.sg)}
\thanks{Y.-D. Yao is with the ECE Department, Stevens Institute of Technology, New Jersey, USA. (Email: yyao@stevens.edu)}

}

\maketitle

\vspace{-0.35 in}

\begin{abstract}
This article investigates the physical-layer security of cognitive radio (CR) networks, which are vulnerable to various newly arising attacks targeting on the weaknesses of CR communications and networking. We first review a range of physical-layer attacks in CR networks, including the primary user emulation, sensing falsification, intelligence compromise, jamming and eavesdropping attacks. Then we focus on the physical-layer security of CR networks against eavesdropping and examine the secrecy performance of cognitive communications in terms of secrecy outage probability. We further consider the use of relays for improving the CR security against eavesdropping and propose an opportunistic relaying scheme, where a relay node that makes CR communications most resistant to eavesdropping is chosen to participate in assisting the transmission from a cognitive source to its destination. It is illustrated that the physical-layer secrecy of CR communications relying on the opportunistic relaying can be significantly improved by increasing the number of relays, showing the security benefit of exploiting relay nodes. Finally, we present some open challenges in the field of relays assisted physical-layer security for CR networks.

\end{abstract}


\IEEEpeerreviewmaketitle

\section{Introduction}

\IEEEPARstart Cognitive radio (CR) [1], [2] emerges as an intelligent radio communications system that is capable of learning its surrounding context and reconfiguring its operating parameters adapted to the time-varying environment. As an enabling technology for spectrum sharing, CR allows an unlicensed user, also called cognitive user (CU), to sense the radio-frequency (RF) environment for detecting whether spectrum bands licensed to primary users (PUs) are occupied by PUs or not [3]. If a licensed band is detected to be unoccupied by PUs, meaning that a spectrum hole is identified, then the CU changes its communications parameters for the sake of transmitting over the detected spectrum hole. Until now, extensive efforts have been devoted to the research and development of CR spectrum sharing systems from different aspects in terms of spectrum sensing, spectrum shaping, spectrum access, and spectrum management [4], [5].

As aforementioned, the physical layer of CR networks is supposed to have the ability of sensing and learning its surrounding RF environment. This, however, is also a critical weakness to be exploited by an adversary for launching malicious activities [6]. For example, the adversary can emit an interfering signal with an intention to modify the actual RF environment, leading legitimate CUs to be misled, compromised and malfunctioned. Also, due to the broadcast nature of radio propagation, any network node within a CU's transmit coverage can overhear the CU's confidential communications and may illegally interpret the confidential information. Therefore, the highly dynamic and open nature of the CR physical layer makes cognitive communications become extremely vulnerable to various malicious activities resulted from both the internal and external attacks.

Recently, the physical-layer security of CR networks has attracted an increasing research attention [7]. Considerable studies have been conducted to protect CR communications against the primary user emulation attack (PUEA) and denial-of-service (DoS) attack. Specifically, a PUEA intends to emulate a PU and transmits a radio signal with the PU's characteristics over a licensed band, misleading that the band is detected to be occupied by the PU and denied to be accessed by legitimate CUs [8]. By contrast, a DoS attacker emits a radio signal (not necessarily with the same characteristics as the PU's signal) to interfere with the signal reception at legitimate CUs for disrupting CR communications services [9], which is also known as a jammer. It needs to be pointed out that both the PUEA and jammer transmit active signals, which may be detected by legitimate CUs so that certain prevention strategies can be adopted.

In addition to the active PUEA and jammer, cognitive transmission is also vulnerable to an eavesdropper, which is a passive attacker and becomes undetectable, since the eavesdropper just overhears and interprets the CR transmission without transmitting any active signals. Generally, cryptographic techniques relying on secret keys are adopted for protecting the transmission confidentiality against eavesdropping, which, however, introduces an additional system complexity resulted from the secret key management. Moreover, the secret key distribution relies upon a trusted infrastructure, which may be unavailable and even compromised in some cases. To this end, physical-layer security is now emerging as a promising paradigm by exploiting physical characteristics of wireless channels to achieve the perfect secrecy against eavesdropping in an information-theoretic sense [10]. This also has a great potential to address the security of CR communications against eavesdropping.

In this article, we are motivated to examine the security of physical-layer communications for CR networks. {{We first present an in-depth overview of CR physical-layer attacks in Section II, including the PUEA, sensing falsification, intelligence compromise, jamming and eavesdropping attacks. Next, we examine the CR physical-layer security in the face of an eavesdropper in Section III and show that increasing the transmit power is not always beneficial in terms of defending against eavesdropping. In Section IV, we propose the employment of opportunistic relaying for protecting the security of CR communications, which is shown to be an effective means, especially with an increasing the number of relays. Finally, we present a range of open challenging issues in Section V, followed by Section VI, where some concluding remarks are provided.}}

\section{Physical-Layer Attacks in CR Networks}
\begin{figure}
  \centering
  {\includegraphics[scale=0.6]{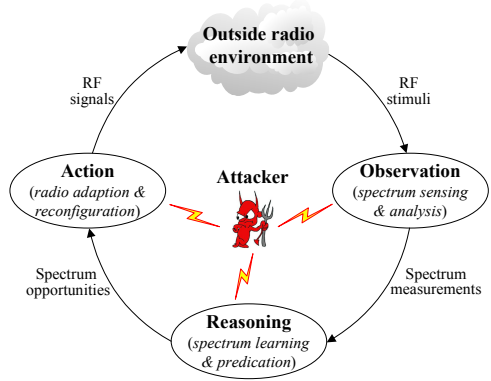}\\
  \caption{Illustration of a typical CR cycle.}\label{Fig1}}
\end{figure}
In this section, we focus on discussing physical-layer attacks in CR networks. As shown in Fig. 1, a CR cycle is comprised of three typical stages, namely the observation, reasoning and action.
Although these three cognitive stages enable a CU to learn its surrounding RF environment and adapt its transmission parameters to any changes in the environment, they are vulnerable to various attacks and introduce additional security threats. Table I summarizes various physical-layer attacks in the observation, reasoning and action phases, including the PUEA, sensing falsification, intelligence compromise, jamming and eavesdropping attacks, which are detailed in the following.

\subsection{PUEA}
{{PUEA refers to an attacker that emulates a PU by transmitting radio signals with the same characteristics as the PU, which prevents legitimate CUs to distinguish the real PU's signal from the PUEA's faked one.
In order to defend against PUEA, a so-called transmitter verification scheme was proposed in [8] by exploiting the location information to verify whether a signal is transmitted from a PU or not. It was assumed in [8] that the PU and PUEA are spatially separated and, moreover, the PU's location is known. However, the location information of PU may be unavailable in some cases. As a consequence, an authentication approach could be employed to differentiate the legitimate PU from PUEA. To be specific, the legitimate PU is registered, whose identity information e.g., the media access control (MAC) address is pre-stored and authenticated. By contrast, the PUEA is typically not registered and its identity remains unknown to legitimate users.}}
\begin{table}
  \centering
  \caption{Summarization of physical-layer attacks in different stages of the CR cycle.}
  {\includegraphics[scale=0.65]{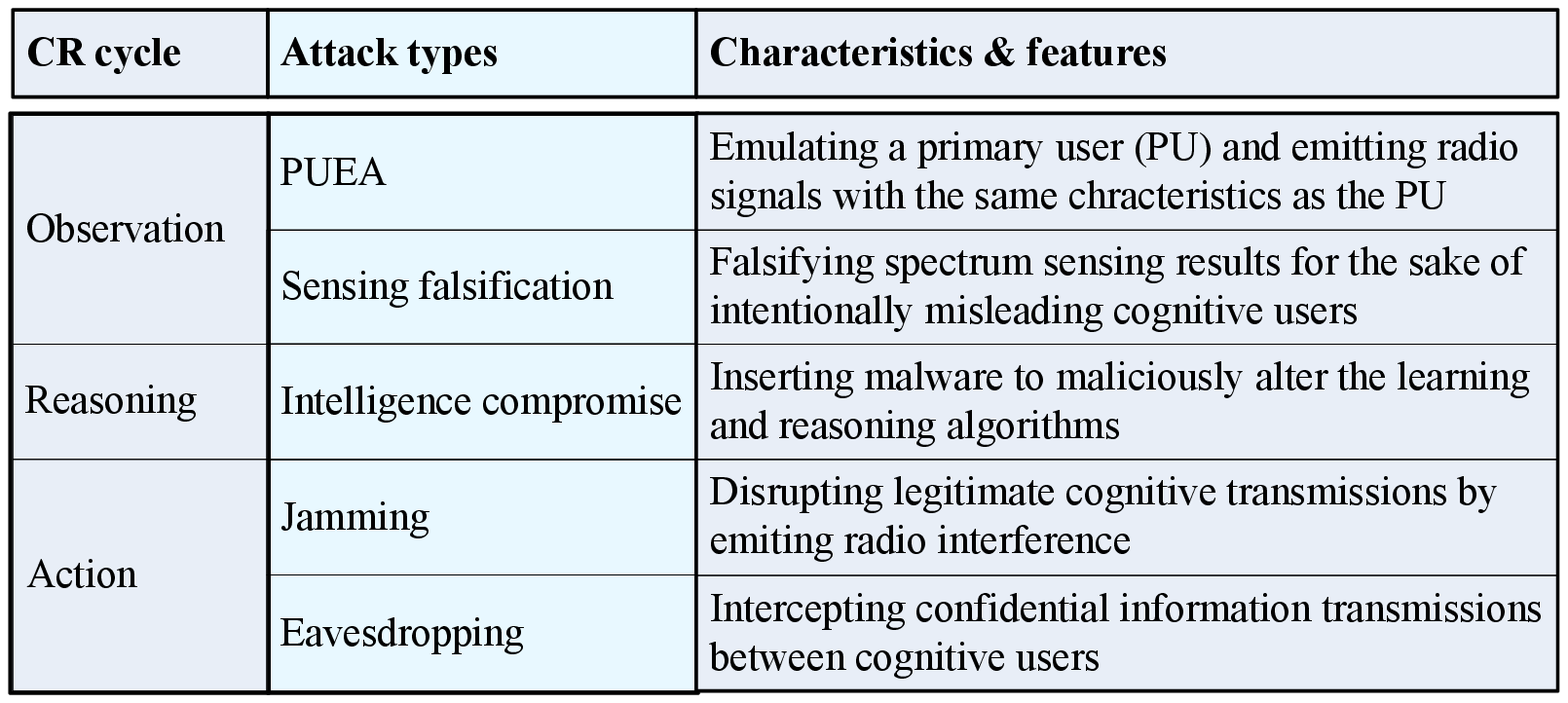}\label{Tab1}}
\end{table}

\subsection{Sensing Falsification}
{{A sensing falsification attacker intends to falsify the spectrum observation and inject its fabricated results to CR networks for the sake of intentionally misleading legitimate CUs.
Typically, the sensing falsification attackers are sparsely distributed and only a small fraction of the total network nodes. Thus, the majority voting is an effective means to mitigate the adverse impact of fabricated observation results on the spectrum sensing performance. As an alternative, a data-cleansing based robust spectrum sensing approach was proposed in [11], where the sparsity of the falsification attack is exploited to effectively filter out the abnormal sensing data. It was shown that the data-cleansing based robust spectrum sensing significantly outperforms conventional spectrum sensing methods in terms of improving the detection probability and false alarm probability in the presence of falsified sensing data.}}


\subsection{Intelligence Compromise}
{{The intelligence compromise is a legitimate CU compromised by an adversary, which maliciously inserts malware into the legitimate CU for the sake of altering its learning and reasoning algorithms, resulting in a negative impact on the node intelligence. An intelligence compromise attacker would inflict damage on the spectrum learning and predication, which may even make the whole CR network become paralyzed.
The intelligence compromise may be just a legitimate CU that is captured and slaved by the adversary, which is thus considered as an inside attacker. Since the intelligence compromised legitimate CU infected by malware still has valid identity, it is difficult to detect and identify the presence of an intelligence compromise attacker. To this end, the automatic code patch is a promising paradigm to protect legitimate CUs against the intelligence compromise, which enables a legitimate CU to be periodically updated. If the code patch fails, it indicates that the legitimate CU may be compromised by an adversary.}}

\subsection{Jamming}
{{A jamming attacker (also known as jammer) attempts to emit a radio signal for interfering with the desired communications between legitimate CUs. As shown in Fig. 1, after identifying an available spectrum opportunity in the observation and reasoning stages, a legitimate CU would be scheduled to transmit its signal to its intended destination over the detected spectrum hole. Due to the broadcast nature of radio propagation, a jammer can easily disrupt the legitimate transmissions between CUs by sending a radio interference with sufficiently high power.
If a jammer is present to interfere with the cognitive transmission, the received signal strength (RSS) and bit error rate (BER) experienced at the desired destination would significantly increase, which can thus be considered as appropriate indicators for detecting the jamming attack. For example, an unusually high RSS (or an excessive BER) may indicate the presence of a jammer. Additionally, spread spectrum is considered as an effective means of defending against jamming attacks. The main spread spectrum techniques include the frequency hopping spread spectrum (FHSS) and direct-sequence spread spectrum (DSSS).}}

\subsection{Eavesdropping}
{{An eavesdropping attacker is to intercept the confidential information transmissions of legitimate CUs. The broadcast nature of wireless propagation makes the cognitive transmissions vulnerable to the eavesdropping attack. When a legitimate CU transmits its data over a detected spectrum hole, any network node within the CU's transmit coverage is capable of overhearing and tapping the CU's transmission. Presently, the cryptography is adopted to protect the communications confidentiality against eavesdropping.
The success of cryptography typically relies on a trusted infrastructure, which, however, may be compromised and becomes untrustworthy [12]. To this end, the information-theoretic security emerges for cognitive radio transmissions by exploiting physical characteristics of wireless channels, referred to as physical-layer security [7], which will be discussed in details in what follows.}}

\section{Physical-Layer Security of Cognitive Radio Communications}
This section presents the physical-layer security of cognitive transmissions from a cognitive source (CS) to its cognitive destination (CD) in the presence of an eavesdropper. As shown in Fig. 2, CS first performs spectrum sensing to detect whether or not a spectrum band is occupied by a primary source (PS) transmitting to its primary destination (PD). If PS is detected to be actively transmitting, CS is not allowed to access the spectrum band for avoiding interfering with the reception of {{PS' signal}}. If PS is detected to be inactive and thus an available spectrum hole is identified, CS would transmit its data to CD over the detected spectrum hole. For notational convenience,
let $P_0$ represent the probability that the spectrum band becomes unoccupied by PS. Additionally, the probability of detection of the presence of PS is denoted by $P_d $, whilst $P_f $ is the probability of false alarm of the presence of PS.
\begin{figure}
  \centering
  {\includegraphics[scale=0.65]{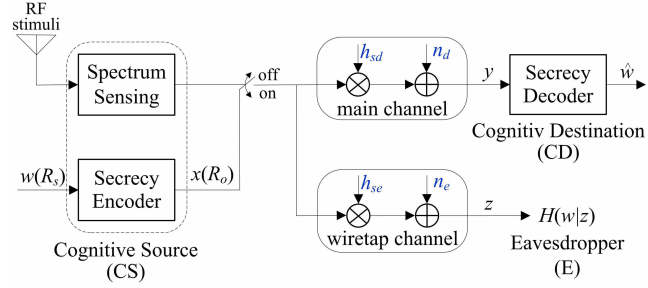}\\
  \caption{A channel model for the secrecy coding based cognitive radio communications.}\label{Fig2}}
\end{figure}

Once a spectrum hole is detected, CS switches on transmitting its confidential data to CD, which may also be overheard by an eavesdropper (E) due to the broadcast nature of radio propagation. {{It is proved in [10] and [15] that when the main channel (from CS to CD) has a better condition than the wiretap channel (from CS to E), physical-layer security can achieve the perfect secrecy against eavesdropping.}} The \emph{secrecy capacity} is shown as the difference between the capacity of the main channel and that of the wiretap channel, which is the maximum rate at which CS can reliably and securely transmit to CD. In order to achieve the secrecy capacity, various secrecy codes (e.g. polar code and lattice code) are devised for practical wireless systems. As shown in Fig. 2, a secrecy encoder (e.g. polar code) encapsulates the CS' confidential data $w$ (with a secrecy rate of $R_s$) {{into}} an overall codeword $x$ (with an increased rate of $R_o$). The rate increase $R_i = R_o - R_s$ represents extra redundancy, which is the cost of providing additional secrecy against eavesdropping. As shown in [12], if the rate cost $R_i$ is higher than the capacity of the wiretap channel, the perfect secrecy can be achieved, i.e., the CS' data transmission is completely secure. Otherwise, the eavesdropper would succeed in intercepting the CS' transmission and a secrecy outage event happens in this case.

Next, CS transmits its codeword $x$ to CD at a power of $P_s$, which is scaled with a wireless fading $h_{sd}$ of the main channel and deteriorated by an additive white Gaussian noise (AWGN) $n_d$. Meanwhile, the codeword transmission is also overheard by E over the wiretap channel, where a wireless fading $h_{se}$ and an AWGN $n_e$ are encountered. Throughout this article, both the main channel and wiretap channel are independent of each other and modeled as Rayleigh fading, implying that $|h_{sd}|^2$ and $|h_{se}|^2$ are independent exponential random variables (RVs) with respective means of $\sigma^2_{sd}$ and $\sigma^2_{se}$. Moreover, the AWGNs received at the CD and E are assumed to be with zero mean and a variance of $N_0$. It is worth mentioning that the miss detection of the presence of PS {{may happen}} due to the background noise, which would cause mutual interference between the primary and cognitive users. To limit the mutual interference level, IEEE 802.22 standard requires $P_d > 0.9$ and $P_f < 0.1$ [2], which is used throughout this article. The transmit power of PS is represented by $P_p$. In addition, fading magnitudes of the wireless channels from PS to CD and E are, respectively, denoted by $|h_{ps}|^2$ and $|h_{pe}|^2$, which are independent exponential RVs with respective means of $\sigma^2_{pd}$ and $\sigma^2_{pe}$.

In order that CS can achieve an ergodic capacity of the main channel, the codeword rate $R_o$ is set to $C_{sd}$ which represents an instantaneous capacity of the CS-CD channel.
Similarly, an instantaneous capacity of the wiretap channel (from CS to E) is denoted by $C_{se}$. As discussed above, a secrecy outage event occurs when the wiretap channel capacity becomes higher than the rate cost $R_i$. It needs to be pointed out that CS starts transmitting its data only when a spectrum hole is detected. Hence, the probability of occurrence of secrecy outage event (called secrecy outage probability) is calculated under the condition that the spectrum band is detected to be unoccupied by PS. Hence, the secrecy outage probability of CS-CD transmissions is given by
\begin{equation}
P_{sout}  = \Pr \left( {C_{se}  > R_i |\hat H_0 } \right) = \Pr \left( {C_{sd}  - C_{se}  < R_s |\hat H_0 } \right),
\end{equation}
where $\hat H_0$ means that the spectrum band is detected idle. In Fig. 3, we show the secrecy outage probability versus signal-to-noise ratio (SNR) $\gamma_s={{P_s }}/{{N_0 }}$ of cognitive radio communications for different secrecy rates with $P_0 = 0.8$, $\gamma_p={{P_p }}/{{N_0 }}=5{\textrm{dB}}$, $\sigma^2_{sd}=1$, $\sigma^2_{pd}=\sigma^2_{pe}=0.2$, and $\sigma^2_{se}=0.1$. {{It needs to be pointed out that the primary and secondary users are spatially separated in two different wireless networks, thus a channel gain between two heterogeneous users from different wireless networks (e.g. $\sigma^2_{pd}$) is set to be smaller than that between two homogeneous users from the same network (e.g. $\sigma^2_{sd}$) [5], [14]. Moreover, following the physical-layer security literature [7], [10] and [15], the wiretap channel is typically assumed to be a degraded version of the main channel, and thus the gain of wiretap channel $\sigma^2_{se}$ is considered to be less than that of the main channel $\sigma^2_{sd}$.}}

\begin{figure}
  \centering
  {\includegraphics[scale=0.55]{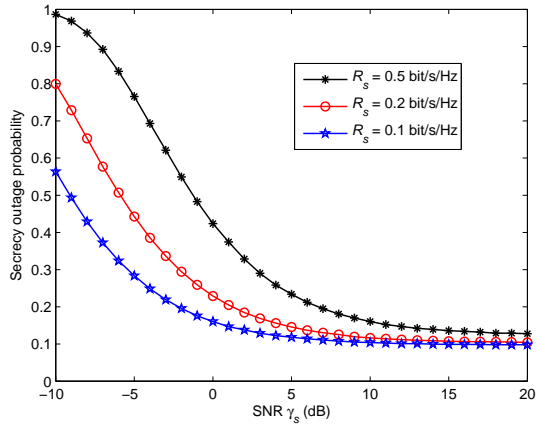}\\
  \caption{Secrecy outage probability versus SNR $\gamma_s$ for different secrecy rates.}\label{Fig3}}
\end{figure}
As shown in Fig. 3, as the secrecy rate increases from $R_s=0.1{\textrm{ bit/s/Hz}}$ to $0.5{\textrm{ bit/s/Hz}}$, the secrecy outage probability of cognitive radio communications increases accordingly. This means that the physical-layer security degrades with an increased rate, showing a tradeoff between the security and throughput. One can also see from Fig. 3 that as the SNR $\gamma_s$ increases, the secrecy outage probability initially decreases and finally converges to a constant value. It implies that a secrecy outage floor happens in high SNR region, which can not be improved by increasing the transmit power. {{This is because that although increasing the transmit power can improve the received signal strength at the legitimate CD, an enhanced signal version is also received at the eavesdropper, which leads to the fact that no secrecy outage improvement is achieved with an increasing transmit power, i.e., a secrecy outage floor occurs in high SNR region.}} We are thus motivated to explore how the secrecy outage floor can be reduced by using e.g. opportunistic relaying, as will be discussed in the following section.

\section{Opportunistic Relaying for Enhancing Physical-Layer Security}
In this section, we examine the employment of opportunistic relaying for the enhancement of physical-layer security in CR networks. As shown in Fig. 4, $N$ relay nodes (RNs) are assumed to be available for assisting the transmission from CS to CD, where the amplify-and-forward (AF) protocol is considered when RNs retransmit the CS' data to CD. To be specific, when a spectrum hole is detected, CS first transmits its signal $x$ to CD, which can be overheard by E and $N$ RNs. In the opportunistic relaying, only a single RN will be chosen among the $N$ RNs to forward an amplified version of its received signal using a scaling factor (without any sort of decoding), which is also overhead by E for interception purposes. In this way, both CD and E can receive two copies of the CS' signal, which are transmitted from the CS and the selected RN, respectively. For simplicity, the selection diversity combining (SDC) method is considered for both the CD and E, meaning that a received signal with higher SNR is adopted for decoding the CS' signal.
\begin{figure}
  \centering
  {\includegraphics[scale=0.6]{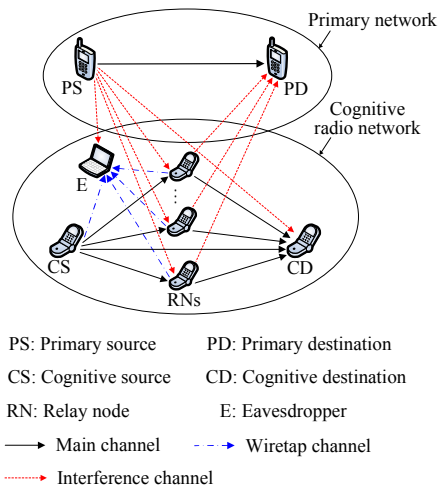}\\
  \caption{A cognitive relay network consists of one CS, one CD and $N$ RNs in the presence of an E.}\label{Fig4}}
\end{figure}

Given $N$ RNs available in CR networks of Fig. 4, the opportunistic relaying chooses the ``best" RN to participate in forwarding the CS' transmission to CD, aiming to maximize the cognitive physical-layer security against eavesdropping. Without loss of generality, we consider that RN$_i$ is selected among $N$ RNs, which first performs a coherent reception of the CS' signal and then forwards its received signal with a scaling factor for normalization. Due to the broadcast nature of radio propagation, both CD and E can receive the RN$_i$'s signal retransmission and the corresponding signal-to-interference-and-noise ratio (SINR) at CD given by
\begin{equation}
\begin{split}
{\textrm{SINR}}_d^i  = \frac{{|h_{si} |^2 |h_{id} |^2 \gamma _s }}{{|h_{id} |^2 (|h_{pi} |^2 \alpha \gamma _p  + 1) + |h_{si} |^2 (|h_{pd} |^2 \alpha \gamma _p  + 1)}}, \\
 \end{split}
\end{equation}
where $h_{si}$, $h_{id}$, $h_{pi}$ and $h_{pd}$ represent the CS-RN$_i$, RN$_i$-CD, PS-RN$_i$ and PS-CD channels, respectively. Moreover, the parameter $\alpha$ is given by $0$, when the spectrum band is idle (i.e., no primary signal is transmitted from PS). By contrast, if the band is currently occupied by PS, then $\alpha$ is set to $1$. Meanwhile, the SINR received at E denoted by ${\textrm{SINR}}_e^i$ can be similarly obtained by replacing $h_{id}$ and $h_{pd}$ in (2) with $h_{ie}$ and $h_{pe}$ which represent the RN$_i$-E and PS-E channels, respectively. In practice, obtaining the eavesdropper's channel state information (CSI) is impossible, since E is passive and typically keeps silent in CR networks. Motivated by this observation, a RN that maximizes the CD's received SINR i.e. {{${\textrm{SINR}}_d^i$}} is generally selected to forward its received signal, yielding the best RN selection criterion as
\begin{equation}
{\textrm{Best RN}} = \arg \mathop {\max }\limits_{i \in {\cal{R}}} {\textrm{SINR}}_d^i ,
\end{equation}
where ${\cal{R}}$ denotes the set of $N$ RNs and {{${\textrm{SINR}}_d^i$ is given by (2)}}. It can be observed from (3) that the CSIs of the CS-RN$_i$, RN$_i$-CD, PS-RN$_i$ and PS-CD channels are required in carrying out the relay selection without needing the eavesdropper's CSI knowledge. {{Moreover, when $\alpha$ is set to $0$, the relay selection criterion as given by (3) degrades to the conventional so-called harmonic mean selection [13]. This is because that $\alpha=0$ implies no mutual interference occurring between the primary and secondary users, thus the cognitive transmission in this case becomes the same as the conventional wireless communications scenario. From (3), the capacity achieved at CD, denoted by $C_{d}$, can be determined by using the SDC to combine the two received signals from the ``best" RN and CS, respectively. Also, the wiretap channel capacity achieved at E, denoted by $C_{e}$, can be similarly obtained. Like (1), the secrecy outage probability of the opportunistic relaying scheme can be obtained by calculating the probability that the difference between $C_d$ and $ C_e$ falls below the secrecy rate $R_s$.}} Additionally, all the CS-CD, CS-RN$_i$, RN$_i$-CD, PS-CD, PS-RN$_i$, PS-E, CS-E, RN$_i$-E channels are modeled as independent Rayleigh fading with respective variances of $\sigma^2_{sd}$, $\sigma^2_{si}$, $\sigma^2_{id}$, $\sigma^2_{pd}$, $\sigma^2_{pi}$, $\sigma^2_{pe}$, $\sigma^2_{se}$, and $\sigma^2_{ie}$.
\begin{figure}
  \centering
  {\includegraphics[scale=0.55]{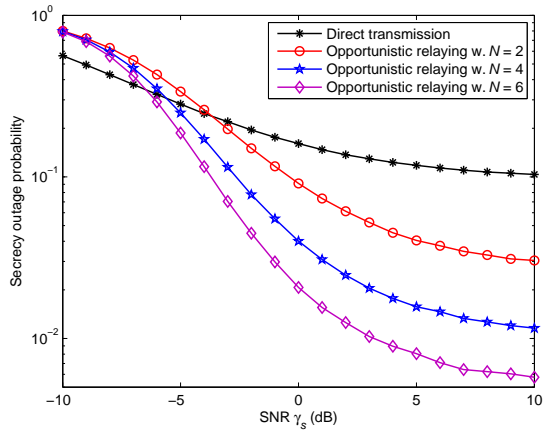}\\
  \caption{Secrecy outage probability comparison between the direct transmission and opportunistic relaying for different number of RNs $N$ with $R_s=0.1{\textrm{bit/s/Hz}}$, $P_0 = 0.8$, $\gamma_p=5{\textrm{dB}}$, $\sigma^2_{sd}=\sigma^2_{si}=\sigma^2_{id}=1$, $\sigma^2_{pd}=\sigma^2_{pi}=\sigma^2_{pe}=0.2$, and $\sigma^2_{se}=\sigma^2_{ie}=0.1$.}\label{Fig5}}
\end{figure}

In Fig. 5, we show the secrecy outage probability versus SNR $\gamma_s$ of the direct transmission (i.e. CS directly transmits to CD without using RNs) and the opportunistic relaying for different number of RNs $N$. As shown in Fig. 5, for all the case of $N=2$, $4$ and $6$, the secrecy outage probability of {{the opportunistic relaying is even worse than that of the direct transmission}} in a low SNR region e.g. $\gamma_s<-6{\textrm{dB}}$. This is because that in the opportunistic relaying scheme, one half of a time slot is wasted by the chosen ``best" RN to retransmit the CS' signal to CD, resulting in a certain loss of the secrecy capacity. {{It is pointed out that although the SDC method is considered at CD for combining its received signals from CS and the ``best" RN, the capacity of the CS-CD channel (i.e. $C_{sd}$) is also scaled by one-half in the opportunistic relaying scheme, since CS transmits only in the first half time slot and remains silent in the second half slot which is occupied by the ``best" RN to retransmit the CS' signal.}} One can observe from Fig. 5 that as the SNR continues increasing, the opportunistic relaying becomes better than the direct transmission in terms of the secrecy outage probability, showing the performance benefit achieved by the proposed opportunistic relaying.

Fig. 5 also shows that with a sufficiently high SNR, the direct transmission and opportunistic relaying schemes converge to their respective secrecy outage floors. Moreover, the secrecy outage floor of the opportunistic relaying is lower than that of the direct transmission. As shown in Fig. 5, as the number of RNs increases from $N=2$ to $6$, the secrecy outage floor of the opportunistic relaying is significantly reduced, showing the physical-layer security advantage of exploiting RNs. {{This is due to the fact that with an increasing number of RNs, it is more likely to choose a RN that can succeed in defending against eavesdropping, thus leading to a reduced secrecy outage floor. Although the opportunistic relaying scheme can effectively protect the wireless transmissions against eavesdropping, it introduces additional system complexity due to the distributed relay management and synchronization. To be specific, multiple RNs are distributed spatially in cognitive radio networks, which need to be effectively managed and synchronized for the sake of performing the opportunistic relay selection. Additionally, in the opportunistic relaying scheme, CD needs to combine its received signals from the ``best" RN and CS, which comes at the cost of extra computational complexity for signal combining.}}

\section{Open Challenges and Future Work}
This section presents some future directions in the research {{field}} of cognitive relay security. Although the opportunistic relaying is shown to enhance the cognitive communications security, there are many challenging issues that still remain open at the time of writing.

\subsection{Joint Relay-and-Jammer Selection}
When CS transmits its signal to CD in the presence of an eavesdropper, a partner node can either be employed as a relay to assist the CS' transmission for enhancing the signal quality received at CD, or act as a jammer to emit artificial noise for contaminating the eavesdropper's signal reception. It is unclear whether it is beneficial to employ the node as a relay (or jammer) in terms of defending the CR communications against eavesdropping. Additionally, given multiple partner nodes available, some nodes may be selected for assisting the CS-CD transmission, while the others may be used as jammers for generating the artificial noise to interfere with the eavesdropper. This is called joint relay-and-jammer selection, which can be considered as a means for improving the cognitive communications security against eavesdropping. Although there are some existing efforts devoted to the joint relay-and-jammer selection, they are limited to the single-relay and single-jammer selection in non-cognitive radio networks. It is of interest to explore a more general framework of multi-relay and multi-jammer selection in cognitive radio networks.

\subsection{Untrusted Relay Detection and Prevention}
As discussed above, the physical-layer security of cognitive radio communications is significantly improved by using the opportunistic relaying in terms of secrecy outage probability. Although the employment of relays is capable of enhancing the security of cognitive communications against eavesdropping, the relays by themselves may not be trusted and attempt to tap the CR communications. For example, if a relay is captured and compromised by an adversary, it becomes untrusted and launches malicious activities (e.g. eavesdropping) in CR networks. It remains unclear about the secrecy performance of CR communications in the face of untrusted relays, which may be considered for future work. Also, it is of high importance to explore the detection and prevention of untrusted relays in CR networks.

\subsection{Field Experiment for Opportunistic Relaying}
IEEE 802.22 is the first worldwide standard designed for the CR based wireless regional area network (WRAN), which enables unlicensed devices to operate in white spaces of the TV broadcast spectrum without causing harmful interference to incumbent users including the TV users and wireless microphones. It is necessary to carry out field experiments for testing the effectiveness of opportunistic relaying in real IEEE 802.22 WRANs in the presence of various attacks. Although the opportunistic relaying is shown to enhance the security of cognitive communications in terms of secrecy outage probability, its security benefit is only proved theoretically based on some simplified assumptions (e.g. perfect CSI knowledge is assumed). It is highly interest to investigate whether the opportunistic relaying is still effective in real WRAN environments in terms of defending against CR attacks.

\section{Conclusion}
In this article, we first presented a comprehensive review on physical-layer attacks in CR networks, including the PUEA, sensing falsification, intelligence compromise, jamming and eavesdropping attacks. The physical-layer security of CR communications in the presence of an eavesdropper was then examined in terms of secrecy outage probability. It was shown that as the transmit power increases, the secrecy outage probability of cognitive communications initially decreases and finally converges to a fixed value, showing that a secrecy outage floor occurs in high SNR regions. In order to improve the physical-layer security of cognitive communications, we considered the use of relays to assist the cognitive communications and proposed an opportunistic relaying scheme. Numerical results showed that upon increasing the number of relays, the opportunistic relaying can significantly reduce the secrecy outage floor of cognitive communications. Additionally, we pointed out some open challenges in the research field of exploiting relays for the physical-layer security of CR networks.

\section{Acknowledgement}
This work was supported by the ``1000 Young Talents Program" of China, the National Natural Science Foundation of China (Grant Nos. 61302104 and 61401223), the Natural Science Foundation of Jiangsu Province (Grant No. BK20140887), and the Scientific Research Foundation of Nanjing University of Posts and Telecommunications (Grant Nos. NY213014 and NY214001).

\end{document}